\documentclass[aps,prc,twocolumn,superscriptaddress,showpacs,floatfix,nofootinbib]{revtex4-1}
\usepackage{amsmath,graphicx,color}

\newcommand{\trento}{T$\mathrel{\protect\raisebox{-2.1pt}{R}}$ENTo}
\begin{document}

\title{Kurtosis of elliptic flow fluctuations}

\author{Rajeev S. Bhalerao}
\affiliation{Department of Physics, Indian Institute of Science Education and Research (IISER), Homi Bhabha Road, Pune 411008, India}
\author{Giuliano Giacalone}
\affiliation{Institut de physique th\'eorique, Universit\'e Paris Saclay, CNRS, CEA, F-91191 Gif-sur-Yvette, France} 
\author{Jean-Yves Ollitrault}
\affiliation{Institut de physique th\'eorique, Universit\'e Paris Saclay, CNRS, CEA, F-91191 Gif-sur-Yvette, France} 
\date{\today}

\begin{abstract}
  Elliptic flow ($v_2$) in ultrarelativistic nucleus-nucleus collisions fluctuates event to event, both in magnitude and in orientation with respect to the reaction plane.
  Even though the reaction plane is not known event to event in experiment, we show that the statistical properties of $v_2$ fluctuations in the reaction plane can be precisely extracted from experimental data.
  Previous studies have shown how to measure the mean, variance and skewness using the first three cumulants $v_2\{2\}$, $v_2\{4\}$ and $v_2\{6\}$. 
  We complement these studies by providing a formula for the kurtosis, which requires an accurate determination of the next cumulant $v_2\{8\}$.
  Using existing data, we show that the kurtosis is positive for most centralities, in contrast with the kurtosis of triangular flow fluctuations, which is negative.  
We argue that these features are robust predictions of fluid-dynamical models. 
\end{abstract}
\maketitle

\section{Introduction}
Anisotropic flow is a spectacular phenomenon observed in ultrarelativistic nucleus-nucleus~\cite{Ackermann:2000tr,ALICE:2011ab,Adare:2011tg,Aidala:2018mcw} and proton-nucleus~\cite{CMS:2012qk,Aad:2012gla} collisions. 
It is understood as resulting from the hydrodynamic response of the quark-gluon plasma to its anisotropic shape at the early stage of the collision~\cite{Ollitrault:1992bk,Alver:2010gr}.
Event to event fluctuations of anisotropic flow~\cite{Alver:2006wh} thus give valuable insight~\cite{Bhalerao:2011yg,Retinskaya:2013gca,Giacalone:2017uqx} into the early-stage dynamics, where the origin of these fluctuations lies~\cite{Albacete:2018bbv}. 
The probability distribution of anisotropic flow has been analyzed in detail~\cite{Aad:2013xma}.
Flow fluctuations are Gaussian to a good approximation~\cite{Voloshin:2007pc}.
Non-Gaussianities are however directly revealed by measurements of higher-order cumulants, such as $v_2\{4\}$ in proton-nucleus collisions~\cite{Aad:2013fja,Chatrchyan:2013nka} or $v_3\{4\}$ in Pb+Pb collisions~\cite{ALICE:2011ab}. 
They are also responsible for the small splitting between $v_2\{4\}$ and $v_2\{6\}$ in Pb+Pb collisions~\cite{Giacalone:2016eyu,Sirunyan:2017fts,Acharya:2018lmh,Mehrabpour:2018kjs}. 
Non-Gaussianities are generic in such microscopic systems, where they appear as corrections to the central limit theorem~\cite{Alver:2008zza,Bhalerao:2011bp,Yan:2013laa}. 
Unlike the situation in the early Universe, where primordial non-Gaussianities are compatible with zero~\cite{Ade:2013ydc} and observed non-Gaussianities are generated during the expansion, the natural expectation in heavy-ion collisions is that non-Gaussianities are already present in the early stages, and partially washed out by the subsequent hydrodynamic expansion~\cite{Giacalone:2016eyu,Abbasi:2017ajp}.
Precise data on non-Gaussian flow fluctuations allow one to test the hydrodynamic picture~\cite{Yan:2013laa,Khachatryan:2015waa} and to constrain models of the initial state~\cite{Giacalone:2017uqx,Gronqvist:2016hym}.

We study fluctuations of elliptic flow, $v_2$, in semi-central nucleus-nucleus collisions, which is the largest and most accurately measured flow phenomenon~\cite{Aamodt:2010pa}. 
Cumulants~\cite{Borghini:2001vi} of the magnitude of $v_2$, denoted by $v_2\{n\}$, have been measured precisely in Pb+Pb collisions for $n=2,4,6,8$~\cite{Aad:2014vba,Sirunyan:2017fts,Acharya:2018lmh}.
It has long been known~\cite{Voloshin:2007pc} that $v_2\{4\}$ is, to a good approximation, the mean $v_2$ projected onto the reaction plane, while the splitting between $v_2\{2\}$ and $v_2\{4\}$ gives access to the variance of the fluctuations.
More recently, it has been shown that the splitting between $v_2\{4\}$ and $v_2\{6\}$ measures the skewness of elliptic flow fluctuations~\cite{Giacalone:2016eyu}.
Here we show that by combining the information from $v_2\{4\}$, $v_2\{6\}$ and $v_2\{8\}$, one can measure the next cumulant, namely, the kurtosis.
The kurtosis of $v_2$ fluctuations in the reaction plane is defined in Sec.~\ref{s:reactionplane}.
We estimate its magnitude and centrality dependence. 
In Sec.~\ref{s:expansion}, we derive a general expression of the kurtosis as a function of the measured cumulants, which is valid for a large system. 
We test the validity of this expression on models of elliptic flow fluctuations. 
In Sec.~\ref{s:data}, we extract the kurtosis from existing data on Pb+Pb collisions at $\sqrt{s_{\rm NN}}=5.02$~TeV. 

\section{Kurtosis of $v_2$ fluctuations in the reaction plane}
\label{s:reactionplane}

\subsection{Definition}

Elliptic flow is the second complex Fourier coefficient of the single-particle distribution $f({\bf p})$~\cite{Luzum:2011mm}:
$V_2=\int e^{2i\varphi} f({\bf p})d{\bf p}/\int f({\bf p})d{\bf p}$, where integration runs over the detector acceptance. 
It can be decomposed into real and imaginary parts: $V_2=v_x+iv_y$. 
In this section, we choose for $\varphi=0$ the direction of impact parameter, or reaction plane.
We focus on the probability distribution of $v_x$, the projection of elliptic flow onto the reaction plane.
This distribution can be characterized by its cumulants. We denote by $\kappa_{n0}$ the cumulant of order $n$.
The origin of this notation will be clarified in Sec.~\ref{s:expansion}. 
The first 4 cumulants are:
\begin{eqnarray}
\label{defkappan0}
\kappa_{10}&=&\langle v_x\rangle,\cr
\kappa_{20}&=&\left\langle( v_x-\langle v_x\rangle)^2\right\rangle,\cr
\kappa_{30}&=&\left\langle( v_x-\langle v_x\rangle)^3\right\rangle,\cr
\kappa_{40}&=&\left\langle( v_x-\langle v_x\rangle)^4\right\rangle-3\kappa_{20}^2, 
\end{eqnarray}
where angular brackets denote an average over many events in a centrality class.
$\kappa_{30}$ and $\kappa_{40}$ vanish if the distribution of $v_x$ is a Gaussian~\cite{Voloshin:2007pc}.

The standardized skewness of the distribution of $v_x$ is defined by 
\begin{equation}
  \label{defgamma1}
\gamma_1\equiv\frac{\kappa_{30}}{(\kappa_{20})^{3/2}}.
\end{equation}
A generic prediction of hydrodynamics is that the distribution of $v_x$ has negative skew, $\gamma_1<0$~\cite{Giacalone:2016eyu}.
The reason is twofold: First, $v_x$ is proportional to the eccentricity in the reaction plane to a good approximation~\cite{Gardim:2011xv,Niemi:2012aj}. Second, the eccentricity is bounded by 1, and this right cutoff skews the distribution to the left.
$\gamma_1$ has been predicted to become more negative as the centrality percentile increases~\cite{Giacalone:2016eyu}. 
Recent experimental analyses~\cite{Sirunyan:2017fts,Acharya:2018lmh} confirm the hydrodynamic prediction: 
$\gamma_1$ reaches the value $-0.4$ at 60\% centrality. 

Our goal in this paper is to extend this analysis to the next cumulant order. The standardized kurtosis of the fluctuations of $v_x$ is defined by
\begin{equation}
  \label{defgamma2}
\gamma_2\equiv\frac{\kappa_{40}}{(\kappa_{20})^2}.
\end{equation}
This quantity, which vanishes if the fluctuations of $v_x$ are Gaussian, is sometimes referred to as ``excess kurtosis'' rather than just ``kurtosis''.
A positive $\gamma_2$ indicates that the distribution has heavier tails than a Gaussian distribution. 

\subsection{Magnitude and centrality dependence}
\label{s:magnitude}

We now investigate the order of magnitude and centrality dependence of the kurtosis $\gamma_2$. 
Abbasi {\it et al.\/}~\cite{Abbasi:2017ajp} have carried out an extensive event-by-event hydrodynamic calculation (14000 events per centrality bin) using Monte Carlo Glauber initial conditions~\cite{Miller:2007ri}.
Their result for the standardized kurtosis $\gamma_2$ (Fig.~3(c) arXiv version v1 of Ref.~\cite{Abbasi:2017ajp}) is essentially compatible with 0 in the 0-60\% centrality range.
They also compute the kurtosis of the initial eccentricity, which is obtained by replacing the elliptic flow $V_2=v_x+iv_y$ with the initial eccentricity  $\varepsilon_2=\varepsilon_x+i\varepsilon_y$ in Eq.~(\ref{defkappan0}).
It is compatible with 0 for central collisions, but increases with the centrality percentile and reaches $0.4$ at 60\% centrality. 
If $V_2$ was proportional to $\varepsilon_2$ in every event~\cite{Gardim:2011xv}, these two quantities would have the exact same $\gamma_2$. 
The fact that $\gamma_2$ is smaller for elliptic flow than for the initial eccentricity is a clear signature of a nonlinear hydrodynamic response. 
A similar phenomenon is observed for the skewness $\gamma_1$, which is reduced by a factor $\simeq 2$ through the hydrodynamic evolution~\cite{Giacalone:2016eyu,Abbasi:2017ajp}.
Interestingly, this reduction is not seen in transport calculations using the AMPT model~\cite{Wei:2018xpm}, where the skewness of elliptic flow is compatible with the skewness of the initial eccentricity.

\begin{figure}[h]
\begin{center}
\includegraphics[width=\linewidth]{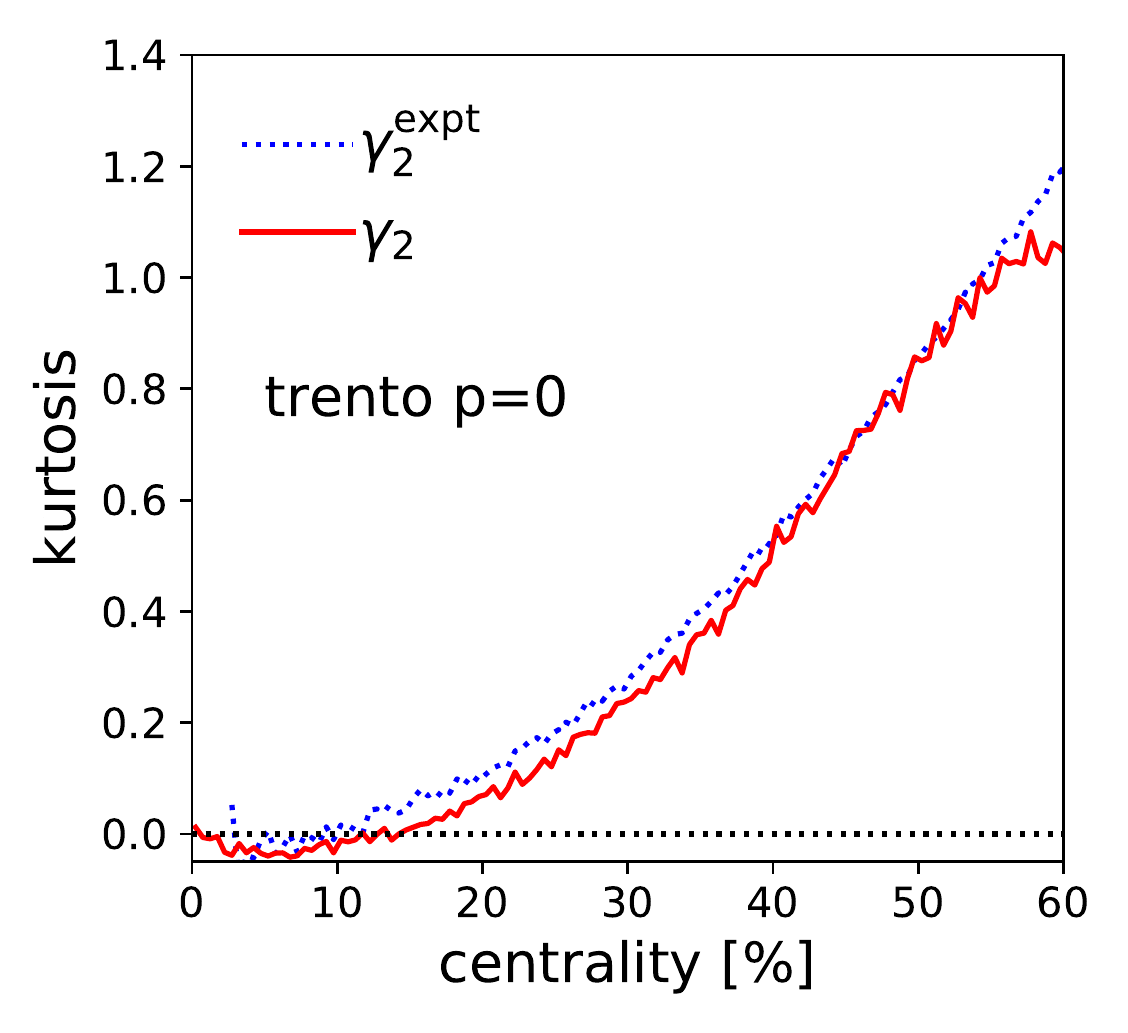} 
\end{center}
\caption{(Color online) 
\label{fig:trento}
Kurtosis of eccentricity fluctuations along the reaction plane in the \trento{} model for Pb+Pb collisions at $\sqrt{s_{\rm NN}}=5.02$~TeV, as a function of centrality percentile. 
Full line: Eq.~(\ref{defgamma2}). Dotted line: Eq.~(\ref{standardized}).
In these equations, we replace the elliptic flow with the initial eccentricity. 
}
\end{figure} 
Even though the hydrodynamic evolution washes out part of the initial non-Gaussianity, it is likely that some of it will remain, which is hidden by statistical errors in the calculation of Ref.~\cite{Abbasi:2017ajp}.
To gain an understanding of the behavior of the kurtosis with the centrality percentile and of its sensitivity to models of initial conditions, we evaluate it within the \trento{} model of initial conditions~\cite{Moreland:2014oya}. 
The \trento{} model has a parameter $p$ which determines how the initial energy density depends on the thickness functions $T_A$ and $T_B$ of colliding nuclei.
We choose the value $p=0$, which corresponds to a density proportional to $\sqrt{T_AT_B}$.
This parametrization reproduces quantitatively the magnitude of anisotropic flow fluctuations in Pb+Pb collisions~\cite{Giacalone:2017uqx}.
We generate $8\times 10^7$ Pb+Pb collisions at $\sqrt{s_{\rm NN}}=5.02$~TeV, which we sort into $0.5\%$ centrality bins. 
The full line in Fig.~\ref{fig:trento} displays $\gamma_2$ of initial eccentricity fluctuations as a function of the centrality percentile. 
It is significantly larger than with the Glauber initial conditions of Ref.~\cite{Abbasi:2017ajp}, and reaches unity at 60\% centrality. 

An interesting feature shown by $\gamma_2$ in Fig.~\ref{fig:trento} is its clear change of sign between 10\% and 15\% centrality.
The negative value of the kurtosis in the 0-10\% centrality range means that the distribution of $\varepsilon_x$ has lighter tails than a Gaussian.
In the context of eccentricity fluctuations, this has been shown to be a consequence of the bound $|\varepsilon_x|<1$ when the distribution of $(\varepsilon_x,\varepsilon_y)$ is azimuthally-symmetric~\cite{Yan:2013laa}, which is the case precisely for the most central collisions.
For this reason, one expects the kurtosis of the initial triangularity, $\varepsilon_3$, and eventually that of triangular flow, $v_3$, to be negative.
The kurtosis of triangular flow fluctuations can be directly obtained from experimental data~\cite{Abbasi:2017ajp}:
\begin{equation}
\label{gamma2v3}
  \gamma_2=-\frac{3}{2}\frac{v_3\{4\}^4}{v_3\{2\}^4},
\end{equation}
where the factor $3/2=\langle\cos^4\phi\rangle/\langle\cos^2\phi\rangle^2$ comes from the projection onto the $x$ axis. 
Thus the observation of a positive $v_3\{4\}$ in Pb+Pb collisions~\cite{ALICE:2011ab} and Xe+Xe collisions~\cite{Giacalone:2018cuy}, in agreement with hydrodynamic predictions~\cite{Giacalone:2017dud}, implies a negative $\gamma_2$, which lies typically between $-0.1$ and $0$~\cite{Abbasi:2017ajp}. 

We now assess the robustness of the \trento{} results in Fig.~\ref{fig:trento} by evaluating the kurtosis for the Elliptic Power distribution~\cite{Yan:2014afa}.
This distribution is the exact~\cite{Gronqvist:2016hym} distribution of the complex eccentricity  $(\varepsilon_x,\varepsilon_y)$ for $N\ge 2$ identical, pointlike sources, randomly distributed in the $(x,y)$ plane with a Gaussian probability distribution.
Its analytic form is
\begin{equation}
p(\varepsilon_x,\varepsilon_y)=
\frac{N-2}{2\pi}
  (1-\varepsilon_0^2)^{\frac{N-1}{2}}\frac{(1-\varepsilon_x^2-\varepsilon_y^2)^{\frac{N}{2}-2}}
{(1-\varepsilon_0\varepsilon_x)^{N-1}},
\label{ellipticpower2d}
\end{equation}
where the parameter $\varepsilon_0$ is the eccentricity of the distribution of the sources.
Eq.~(\ref{ellipticpower2d}) provides a reasonable fit of most models of initial conditions for all centralities.
The cumulants $\kappa_{n0}$ of this distribution can be evaluated analytically in terms of hypergeometric functions~\cite{Abbasi:2017ajp}.
Their asymptotic values for large $N$ are:
\begin{eqnarray}
\label{kappan0EP}
\kappa_{10}&=&\varepsilon_0+{\cal O}\left(\frac{1}{N}\right),\cr
\kappa_{20}&=&\frac{(1-\varepsilon_0^2)^2}{N}+{\cal O}\left(\frac{1}{N^2}\right),\cr
\kappa_{30}&=&-\frac{6\varepsilon_0(1-\varepsilon_0^2)^3}{N^2}+{\cal O}\left(\frac{1}{N^3}\right),\cr
\kappa_{40}&=&\frac{6(12\varepsilon_0^2-1)(1-\varepsilon_0^2)^4}{N^3}+{\cal O}\left(\frac{1}{N^4}\right).
\end{eqnarray}
The cumulants decrease by successive powers of $1/N$ as the order increases.
The skewness and kurtosis are given by: 
\begin{eqnarray}
\label{skewkurtEP}
\gamma_1&=&-\frac{6\varepsilon_0}{\sqrt{N}}+{\cal O}\left(\frac{1}{N^{3/2}}\right),\cr\cr
\gamma_2&=&\frac{6\left(12\varepsilon_0^2-1\right)}{N}+{\cal O}\left(\frac{1}{N^2}\right). 
\end{eqnarray}
They vanish in the limit $N\to\infty$, as expected from the central limit theorem. 
The skewness $\gamma_1$ is negative. 
The sign of $\gamma_2$ is driven by the mean eccentricity $\varepsilon_0$: it is negative for $\varepsilon_0<0.28$ and positive for $\varepsilon_0>0.29$.
Therefore, the observation that $\gamma_2$ is larger in our \trento{} calculation than in the Glauber model~\cite{Abbasi:2017ajp} seems naturally explained by the fact that the \trento{} model presents a larger eccentricity in the reaction plane~\cite{Yan:2014nsa}.

\begin{figure*}[t!]
\begin{center}
\includegraphics[width=.47\linewidth]{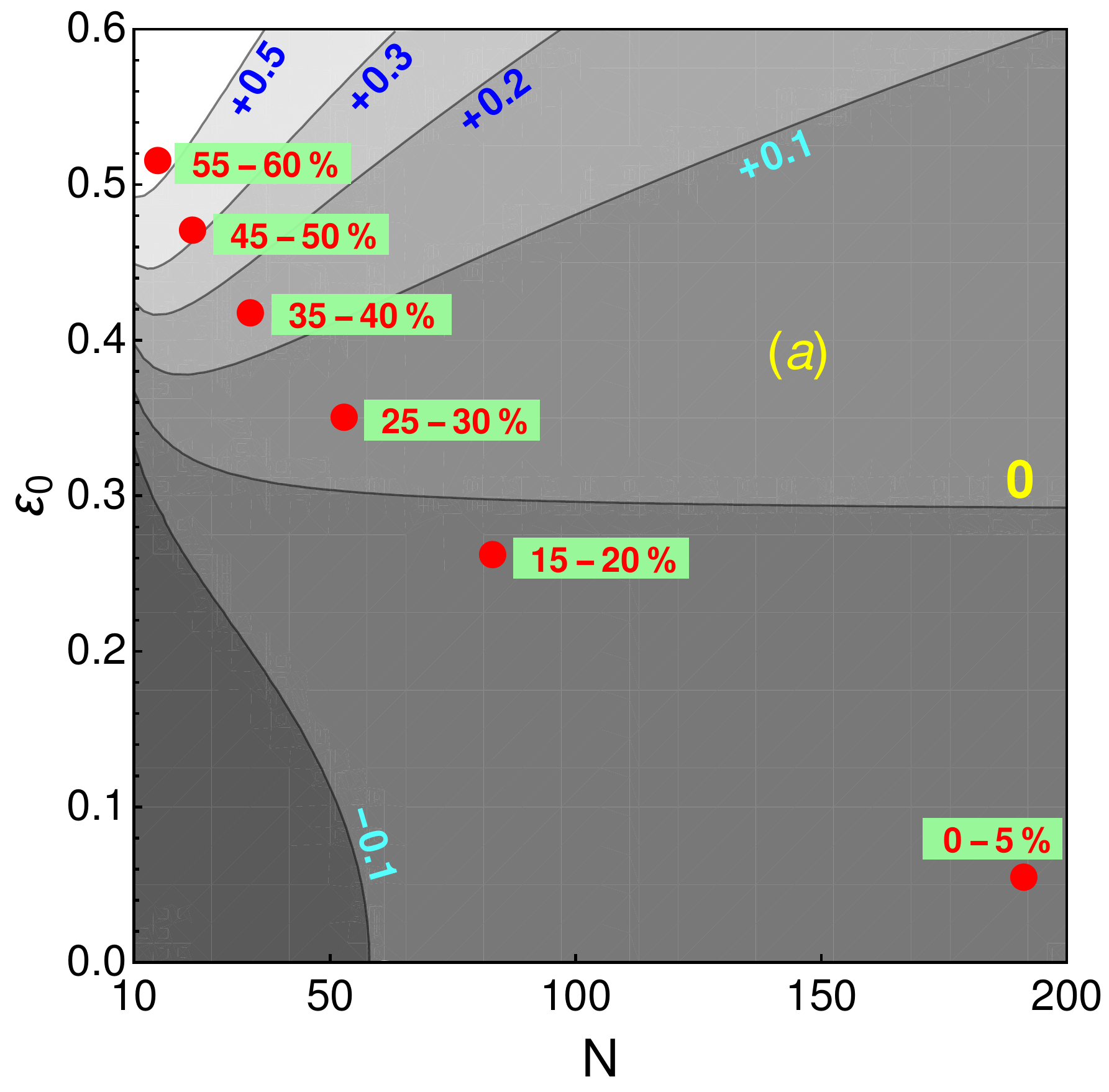}\hspace{25pt} \includegraphics[width=.47\linewidth]{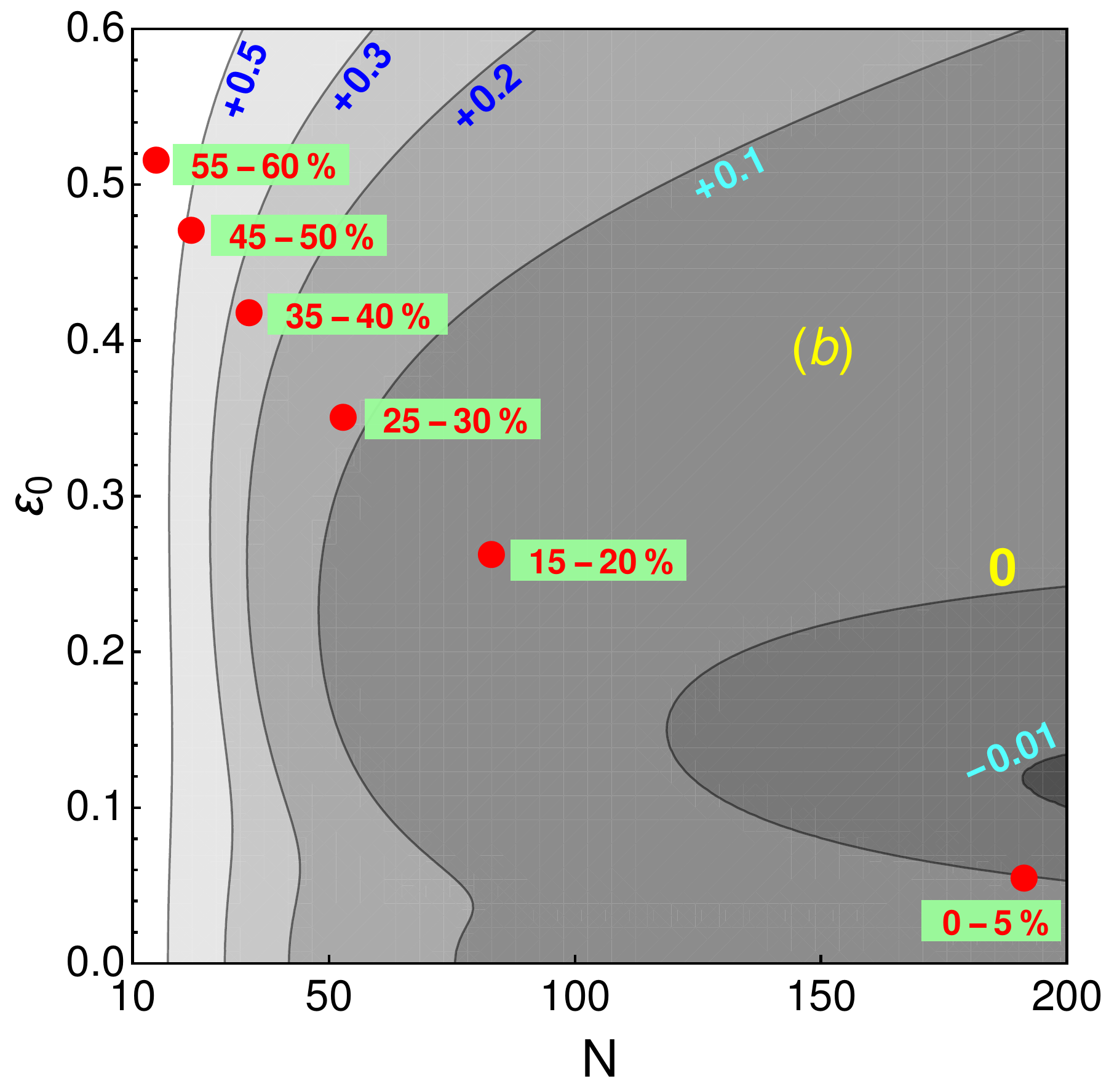}
\end{center}
\caption{(Color online) 
\label{fig:ellpower}
Contour plots of the standardized kurtosis, $\gamma_2$, defined in Eq.~(\ref{defgamma2}) [panel (a)], and of the estimated kurtosis, $\gamma_2^{\rm expt}$, defined by Eq.~(\ref{standardized}) [panel (b)], computed using the Elliptic Power distribution.
The quantities $\varepsilon_0$ and $N$ are the eccentricity of the sources, and their number, respectively, as in Eq.~(\ref{ellipticpower2d}).
Symbols correspond to Elliptic Power fits to the \trento{} calculations (see Fig.~\ref{fig:trento}) in a few centrality windows. 
}
\end{figure*}

In order to check the validity of the asymptotic result (\ref{skewkurtEP}), we evaluate $\gamma_2$ numerically for realistic values of $N$.
Figure~\ref{fig:ellpower} (a) displays contour plots of $\gamma_2$ in the $(N,\varepsilon_0)$ plane.
In order to compare with the values of $\gamma_2$ from the \trento{} calculation, displayed in Fig.~\ref{fig:trento}, we have fitted the eccentricity distributions from the \trento{} model with Eq.~(\ref{ellipticpower2d}) in a few centrality intervals.
The resulting values of $(N,\varepsilon_0)$ are displayed as symbols in Fig.~\ref{fig:ellpower}.
The values of $\gamma_2$ in Figs.~\ref{fig:trento} and \ref{fig:ellpower} (a) are not in quantitative agreement, which means that the fit of \trento{} with the Elliptic Power distribution is not perfect.
However, the order of magnitude and change of sign are reproduced, and seem to be robust predictions of hydrodynamics. 

In conclusion, the positive sign of the kurtosis in non-central collisions appears to be a generic consequence of a large eccentricity in the reaction plane.

\section{Expressing the kurtosis as a function of measured cumulants}
\label{s:expansion}

We now explain how the kurtosis $\gamma_2$ can be extracted from experiment.
This is not trivial because experiments measure the magnitude $v_2\equiv\sqrt{v_x^2+v_y^2}$, not $v_x$ and $v_y$ separately.
In this Section, we show that one can however reconstruct $\gamma_2$ for a large system.\footnote{By large system, we mean a nucleus-nucleus collision, and we have tested that our formalism works up to 60\% centrality.}
We first define the cumulants in two coordinate systems:
(1) The reaction plane coordinate system, where $x$ is the direction of impact parameter, or reaction plane, as in the previous section. 
(2) The detector system, where $x$ denotes a fixed orientation with respect to the detector. This is the natural coordinate system for experiments. 
The usual cumulant of order $n$ measured in experiments, denoted by $v_2\{n\}$~\cite{Borghini:2001vi}, is defined in this system.

\subsection{Cumulants in the reaction plane coordinate system} 

Cumulants are a double sequence $\kappa_{n_x,n_y}$ with $n_x,n_y\ge 0$, which completely specify the probability distribution of $(v_x,v_y)$. They are defined by~\cite{Abbasi:2017ajp}:
\begin{equation}
  \label{generating}
  \ln\left\langle e^{k_xv_x+k_yv_y}\right\rangle=
  \sum_{n_x,n_y}\frac{k_x^{n_x}}{n_x!}\frac{k_y^{n_y}}{n_y!}\kappa_{n_x,n_y}.
\end{equation}
The probability distribution of  $(v_x,v_y)$ is symmetric with respect to the reaction plane in the absence of parity violation~\cite{Kharzeev:2004ey,Voloshin:2004vk}.
This implies that it is an even function of $v_y$, so that the only nonvanishing cumulants are those with even $n_y$.
Let us write the expansion explicitly, keeping all terms with $n_x+n_y\le 4$:
\begin{eqnarray}
  \label{generating4}
  \ln\left\langle e^{k_xv_x+k_yv_y}\right\rangle&=&
  k_x\kappa_{10}\cr
  &+&  \frac{k_x^2}{2}\kappa_{20}+\frac{k_y^2}{2}\kappa_{02}\cr
  &+&  \frac{k_x^3}{6}\kappa_{30}+\frac{k_xk_y^2}{2}\kappa_{12}\cr
  &+&  \frac{k_x^4}{24}\kappa_{40}+ \frac{k_y^4}{24}\kappa_{04}+
  \frac{k_x^2k_y^2}{4}\kappa_{22}. 
\end{eqnarray}
The four lines in the right-hand side correspond respectively to the mean, variance, skewness and kurtosis of the distribution of $(v_x,v_y)$.
The cumulants $\kappa_{n0}$ coincide with those defined in Eq.~(\ref{defkappan0}). 

\subsection{Measured cumulants}

We now recall the definition of the measured cumulants $v_2\{n\}$.
The only difference with the previously defined cumulants is the coordinate system.
We start again from the generating function Eq.~(\ref{generating}), where the $x$ axis now denotes a fixed direction in the detector. 
We first evaluate this generating function in the simple case where $v_2$ is the same for all events, but the orientation of the reaction plane $\Phi_R$ is random.
We write  $v_x=v_2\cos 2\Phi_R$, $v_y=v_2\sin 2\Phi_R$.
Averaging over events amounts to averaging over $\Phi_R$ if the detector is azimuthally symmetric:
\begin{equation}
\label{constantv2}
  \ln\left\langle e^{k_xv_x+k_yv_y}\right\rangle=
\ln I_0(kv_2)=\sum_{n=2}^{\infty} c_n k^n (v_2)^n,
\end{equation}
where $I_0$ is the modified Bessel function of the first kind, $k\equiv \sqrt{k_x^2+k_y^2}$, and $c_n$ are rational coefficients which vanish for odd $n$: $c_2=1/4$, $c_4=-1/64$, $c_6=1/576$, $c_8=-11/49152$...  

In the general case where $v_2$ fluctuates event to event, $\langle e^{k_xv_x+k_yv_y}\rangle$ depends only on $k^2$ by azimuthal symmetry.
One can write, without any loss of generality, 
\begin{equation}
\label{defv2n}
  \ln\left\langle e^{k_xv_x+k_yv_y}\right\rangle=\ln \langle I_0(kv_2)\rangle=
  \sum_{n=2}^{\infty}c_n k^nv_2\{n\}^n,
\end{equation}
where the coefficients $c_n$ are defined by Eq.~(\ref{constantv2}).
Eq.~(\ref{defv2n}) defines $v_2\{n\}^n$ for all even $n$.
Comparison with Eq.~(\ref{constantv2}) shows that if $v_2$ is the same for all events, then, $v_2\{n\}$ coincides with $v_2$ for all $n$.
However, despite this (perhaps unfortunate) notation, $v_2\{n\}^n$ defined by Eq.~(\ref{defv2n}) can have positive or negative sign if $v_2$ fluctuates event to event. 

\subsection{Conversion from one coordinate system to the other}

One can express $v_2\{n\}^n$  as a function of the $\kappa_{n_xn_y}$ in Eq.~(\ref{generating}) systematically in the following way:
\begin{itemize}
\item{Write $k_x=k\cos\theta$, $k_y=k\sin\theta$ in Eq.~(\ref{generating}).}
\item{Exponentiate Eq.~(\ref{generating}) and average over $\theta$.}
\item{Take the logarithm, expand in powers of $k$, and match the result order by order to the right-hand side of Eq.~(\ref{defv2n}).}
\end{itemize}
As an illustration, one finds the following exact expressions for the first three cumulants~\cite{Abbasi:2017ajp}:
\begin{eqnarray}
  \label{exact}
v_2\{2\}^2&=&\kappa_{10}^2\cr&&+\kappa_{20}+\kappa_{02},\cr
v_2\{4\}^4&=&\kappa_{10}^4\cr
&&+2 \kappa_{10}^2(\kappa_{02}-\kappa_{20})\cr
&&-4 \kappa_{10}(\kappa_{30}+\kappa_{12})- (\kappa_{20}-\kappa_{02})^2\cr
&& - (\kappa_{04}+\kappa_{40}+2\kappa_{22}),\cr
v_2\{6\}^6 &=& \kappa_{10}^6 \cr
&&+ 3\kappa_{10}^4(\kappa_{02}-\kappa_{20})\cr
&&-2\kappa_{10}^3(2\kappa_{30}+3\kappa_{12})\cr
&&+\frac{3}{2}\kappa_{10}^2(\kappa_{40}-\kappa_{04})- 6\kappa_{10}\kappa_{30}(\kappa_{02}-\kappa_{20})\cr
&&+\frac{3}{2}(\kappa_{04}-\kappa_{40})(\kappa_{02}-\kappa_{20})
+\frac{5}{2} \kappa_{30}^2+ 3\kappa_{30}\kappa_{12}\cr &&+\frac{9}{2} \kappa_{12}^2+\frac{3}{2}\kappa_{10}(\kappa_{50}+\kappa_{14}+2\kappa_{32})\cr
&&+\frac{3}{4}(\kappa_{24}+\kappa_{42})+\frac{\kappa_{60}}{4}+\frac{\kappa_{06}}{4}.
\end{eqnarray}

One cannot invert these relations and reconstruct the double sequence $\kappa_{n_xn_y}$ from the single sequence $v_2\{n\}$. 
Simplifications occur, however, for a large system.
Inspired by the Elliptic Power distribution, where $\kappa_{n_xn_y}$ is of order $N^{1-n_x-n_y}$, we expand Eqs.~(\ref{exact}) in powers of $1/N$.
Doing so, one finds that, to leading order in $1/N$, the differences between successive cumulants only involve the single sequence $\kappa_{n0}$:
\begin{align}
\label{splittings}
\nonumber v_2\{2\}^2-v_2\{4\}^2&=2\kappa_{20}+{\cal O}\left(\frac{1}{N^2}\right),\\
\nonumber v_2\{4\}^3-v_2\{6\}^3&=-\kappa_{30}+{\cal O}\left(\frac{1}{N^3}\right),\\
v_2\{4\}^4-12v_2\{6\}^4+11v_2\{8\}^4&=-\frac{8}{3}\kappa_{40}+{\cal O}\left(\frac{1}{N^4}\right).
\end{align}
Thus the splitting between order 2 and order 4 is due to elliptic flow fluctuations, as has long been known~\cite{Bhalerao:2006tp}.
The splitting between 4 and 6 is the skewness~\cite{Giacalone:2016eyu}.
The new result is the last line of Eq.~(\ref{splittings}), which shows that the kurtosis $\kappa_{40}$ can be extracted by combining orders 4, 6, and 8, in a way that eliminates the contribution from the skewness $\kappa_{30}$.

The observation that to leading order in $1/N$, the splittings between successive cumulants only involve $\kappa_{n0}$, that is, fluctuations projected onto the reaction plane, can be understood as follows.
Elliptic flow in a given event can be decomposed as $v_2=\sqrt{(\kappa_{10}+\delta_x)^2+\delta_y^2}$, where $\kappa_{10}$ is the mean elliptic flow in the reaction plane and $\delta_x$ and $\delta_y$ denote the fluctuations in the directions parallel and perpendicular to the reaction plane.
Expanding in powers of the fluctuation, one finds to leading order
\begin{equation}
  v_2\simeq\kappa_{10}+\delta_x+\frac{\delta_y^2}{2\kappa_{10}}.
\end{equation}
Thus, fluctuations along the $y$ direction enter at higher order than fluctuations along the $x$ direction. 

If one keeps only the leading term in the right-hand side of Eqs.~(\ref{splittings}), one obtains approximate expressions of the standardized skewness and kurtosis of $v_x$ fluctuations, defined by Eqs.~(\ref{defgamma1}) and (\ref{defgamma2}), in terms of measured quantities:
\begin{eqnarray}
\label{standardized}
  \gamma_1&\simeq &\gamma_1^{\rm expt}\equiv-2^{3/2}\frac{v_2\{4\}^3-v_2\{6\}^3}
  {\left(v_2\{2\}^2-v_2\{4\}^2\right)^{3/2}},\cr
  \gamma_2&\simeq &\gamma_2^{\rm expt}\equiv-\frac{3}{2}\frac{v_2\{4\}^4-12v_2\{6\}^4+11v_2\{8\}^4}{\left(v_2\{2\}^2-v_2\{4\}^2\right)^2}.
\end{eqnarray} 
The expression of the skewness is essentially equivalent to that proposed in Ref.~\cite{Giacalone:2016eyu}, but slightly simpler. 
The expression of the kurtosis is new, but 
similar expressions have been derived by considering fluctuations of the flow magnitude alone~\cite{Jia:2014pza}.
$\gamma_2^{\rm expt}$ vanishes if
\begin{equation}
  \label{v68splitting}
  v_2\{6\}-v_2\{8\}=\frac{1}{11}\left(v_2\{4\}-v_2\{6\}\right),
\end{equation}
where we have linearized Eq.~(\ref{standardized}) by taking into account the observation that the splittings are very small in practice.
The ALICE collaboration~\cite{Acharya:2018lmh} has found that data satisfy Eq.~(\ref{v68splitting}) within error bars.
Measuring a non-trivial value of the kurtosis requires very precise data, as we shall see in Sec.~\ref{s:data}.

\subsection{Numerical tests}

We now check that $\gamma_2^{\rm expt}$ defined by Eq.~(\ref{standardized}) provides a reasonable approximation of the kurtosis $\gamma_2$ defined by Eq.~(\ref{defgamma2}) for heavy-ion collisions. 
We first use the \trento{} simulation. 
The dotted line  in Fig.~\ref{fig:trento} is $\gamma_2^{\rm expt}$ defined by Eq.~(\ref{standardized}). 
For all centralities, we find that the absolute difference between the two estimates of the kurtosis is of order 0.01.
Therefore, $\gamma_2^{\rm expt}$ is an excellent approximation of $\gamma_2$ in noncentral collisions, where the kurtosis is positive and of order 0.1.

An independent test of Eq.~(\ref{standardized}) is provided by the Elliptic Power distribution, which is a toy model in which one can evaluate both $\gamma_2$ and $\gamma_2^{\rm expt}$. 
Fig.~\ref{fig:ellpower} (b) displays contour plots of  $\gamma_2^{\rm expt}$ in the $(N,\varepsilon_0)$ plane. 
If Eq.~(\ref{standardized}) was exact, panels (a) and (b) would be identical.
On the contrary, the two panels look very different at first sight.
This is not surprising, as the approximate equality in Eq.~(\ref{standardized}) only holds if fluctuations are small corrections to the mean eccentricity. 
In other terms, it is valid only if both $N$ and $\varepsilon_0$ are large enough. 
A closer examination of Fig.~\ref{fig:ellpower} indeed confirms that agreement between panels (a) and (b) becomes better as one moves to the upper part (large $\varepsilon_0$) and to the right (large $N$) of the figure.
As in the case of the \trento{} simulation, the difference $\gamma_2^{\rm expt}-\gamma_2$ is everywhere positive.
This difference is actually larger for the Elliptic Power distribution than for the \trento{} simulation, so that the good agreement seen in Fig.~\ref{fig:trento} might be a lucky coincidence.
However, even for the Elliptic Power distribution, $\gamma_2^{\rm expt}$ remains a reasonable approximation of $\gamma_2$ for semi-central nucleus-nucleus collisions.

\section{Experimental data}
\label{s:data}

\begin{figure}[t!]
\begin{center}
\includegraphics[width=\linewidth]{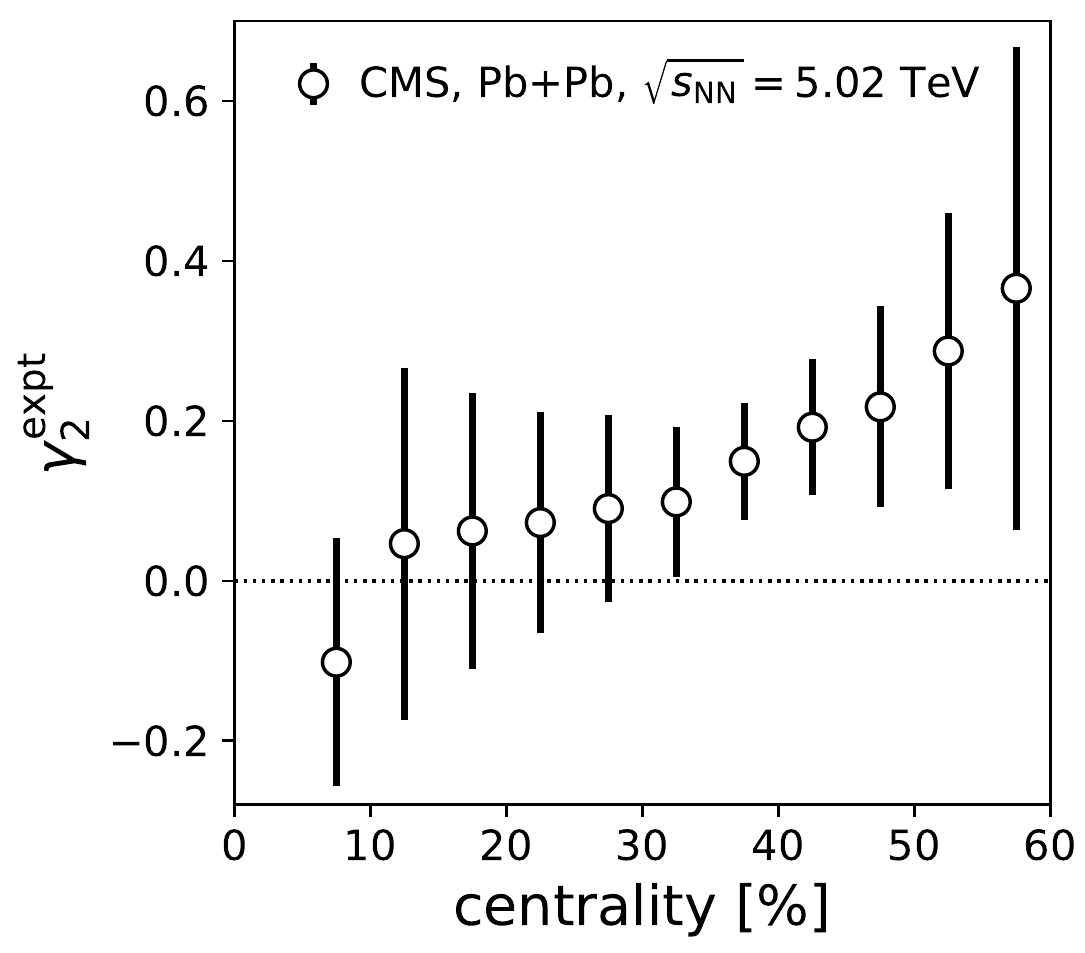}
\end{center}
\caption{(Color online) 
\label{fig:data}
Kurtosis of $v_2$ fluctuations in Pb+Pb collisions at $\sqrt{s_{_\text{NN}}} = 5.02$~TeV estimated using Eq.~(\ref{standardized}) and CMS data~\cite{Sirunyan:2017fts}. 
}
\end{figure} 

Finally, we estimate the kurtosis in Pb+Pb collisions by applying Eq.~(\ref{standardized}) to CMS data~\cite{Sirunyan:2017fts}.
The result is displayed in Fig.~\ref{fig:data}. 
Error bars have been estimated by assuming that the errors on the ratios $v_2\{8\}/v_2\{6\}$ and $v_2\{6\}/v_2\{4\}$ are independent.
Despite the large error bars, there is clear evidence that the kurtosis is positive above 30\% centrality, in agreement with the theoretical calculations of Figs.~\ref{fig:trento} and \ref{fig:ellpower}. 
We have also extracted the kurtosis from ALICE data~\cite{Acharya:2018lmh}.
The result (not shown) is compatible with CMS data, but with much larger error bars, and does not show any evidence of a non-zero kurtosis.
Note that the results of Abbasi {\it et al.\/}~\cite{Abbasi:2017ajp} for the kurtosis of Monte Carlo Glauber initial conditions are of the same magnitude as the experimental data in Fig.~\ref{fig:data}.
However, this kurtosis is washed out by the hydrodynamic evolution, which implies that the kurtosis of the Glauber model is too small. 

We deem that hydrodynamic calculations with extensive statistics, like those of Ref.~\cite{Abbasi:2017ajp}, should be carried out with initial conditions displaying a larger initial kurtosis, such as the \trento{} $p=0$ model shown in Fig.~\ref{fig:trento}.

\section{Conclusions}

We have argued that the kurtosis of elliptic flow fluctuations along the direction of impact parameter is positive in non-central Pb+Pb collisions.
This is a notable difference with respect to the kurtosis of triangular flow fluctuations, which is observed to be negative in experimental data, as predicted by hydrodynamic calculations.
If elliptic flow was a pure linear response to the initial eccentricity, the kurtosis of elliptic flow fluctuations would be equal to the kurtosis of initial eccentricity fluctuations.
However, state-of-the-art hydrodynamic calculations predict that the non-Gaussianities (skewness and kurtosis) are significantly reduced by the hydrodynamic evolution, so that these observables probe hydrodynamics in the nonlinear regime. 


We have provided a formula to extract the kurtosis of elliptic flow fluctuations from high-precision measurements of $v_2\{2\}$, $v_2\{4\}$, $v_2\{6\}$ and $v_2\{8\}$.
This requires very high statistics of events, and, at the present moment, only Run2 CMS data allow for the extraction of a significant $\gamma_2$, which indeed turns out to be positive above 30\% centrality.
The usefulness of investigating flow fluctuations at such a detailed level is nevertheless evident already after our crude extraction: Combined with the precise calculations of Ref.~\cite{Abbasi:2017ajp}, our result provides clear indication that only models displaying large eccentricity, and these are typically the models inspired by high-energy QCD~\cite{Hirano:2005xf,Schenke:2012wb,Niemi:2015qia,Nagle:2018ybc}, have the potential of yielding a $\gamma_2$ of order 0.5 after the hydrodynamic evolution.

The bottom line is that with great precision comes great discriminating power, although at the cost of increasing the statistics of hydrodynamic calculations.
With the advent of LHC3 data, we expect future dedicated analyses using robust methods~\cite{DiFrancesco:2016srj} to characterize the details of flow fluctuations, such as the kurtosis studied in this paper, with unprecedented accuracy.
This will lead to novel insightful tests of the hydrodynamic picture, in the nontrivial regime where the hydrodynamic response driving anisotropic flow is strongly nonlinear.

\section*{Acknowledgments}

RSB would like to acknowledge the hospitality of the IPhT, Saclay,
France where a part of this work was done and the support of the CNRS
LIA (Laboratoire International Associ\'{e}) THEP (Theoretical High
Energy Physics) and the INFRE-HEPNET (IndoFrench Network on High
Energy Physics) of CEFIPRA/IFCPAR (Indo-French Center for the
Promotion of Advanced Research). RSB also acknowledges the support of
the Department of Atomic Energy, India for the award of the Raja Ramanna
Fellowship.


\begin{thebibliography}{99}
\bibitem{Ackermann:2000tr} 
  K.~H.~Ackermann {\it et al.} [STAR Collaboration],
  Phys.\ Rev.\ Lett.\  {\bf 86}, 402 (2001)
  doi:10.1103/PhysRevLett.86.402
  [nucl-ex/0009011].



\bibitem{ALICE:2011ab} 
  K.~Aamodt {\it et al.} [ALICE Collaboration],
  Phys.\ Rev.\ Lett.\  {\bf 107}, 032301 (2011)
  doi:10.1103/PhysRevLett.107.032301
  [arXiv:1105.3865 [nucl-ex]].



\bibitem{Adare:2011tg} 
  A.~Adare {\it et al.} [PHENIX Collaboration],
  Phys.\ Rev.\ Lett.\  {\bf 107}, 252301 (2011)
  doi:10.1103/PhysRevLett.107.252301
  [arXiv:1105.3928 [nucl-ex]].



\bibitem{Aidala:2018mcw} 
  C.~Aidala {\it et al.} [PHENIX Collaboration],
  arXiv:1805.02973 [nucl-ex].



\bibitem{CMS:2012qk} 
  S.~Chatrchyan {\it et al.} [CMS Collaboration],
  Phys.\ Lett.\ B {\bf 718}, 795 (2013)
  doi:10.1016/j.physletb.2012.11.025
  [arXiv:1210.5482 [nucl-ex]].



\bibitem{Aad:2012gla} 
  G.~Aad {\it et al.} [ATLAS Collaboration],
  Phys.\ Rev.\ Lett.\  {\bf 110}, no. 18, 182302 (2013)
  doi:10.1103/PhysRevLett.110.182302
  [arXiv:1212.5198 [hep-ex]].



\bibitem{Ollitrault:1992bk} 
  J.~Y.~Ollitrault,
  Phys.\ Rev.\ D {\bf 46}, 229 (1992).
  doi:10.1103/PhysRevD.46.229



\bibitem{Alver:2010gr} 
  B.~Alver and G.~Roland,
  Phys.\ Rev.\ C {\bf 81}, 054905 (2010)
  Erratum: [Phys.\ Rev.\ C {\bf 82}, 039903 (2010)]
  doi:10.1103/PhysRevC.82.039903, 10.1103/PhysRevC.81.054905
  [arXiv:1003.0194 [nucl-th]].



\bibitem{Alver:2006wh} 
  B.~Alver {\it et al.} [PHOBOS Collaboration],
  Phys.\ Rev.\ Lett.\  {\bf 98}, 242302 (2007)
  doi:10.1103/PhysRevLett.98.242302
  [nucl-ex/0610037].



\bibitem{Bhalerao:2011yg} 
  R.~S.~Bhalerao, M.~Luzum and J.~Y.~Ollitrault,
  Phys.\ Rev.\ C {\bf 84}, 034910 (2011)
  doi:10.1103/PhysRevC.84.034910
  [arXiv:1104.4740 [nucl-th]].



\bibitem{Retinskaya:2013gca} 
  E.~Retinskaya, M.~Luzum and J.~Y.~Ollitrault,
  Phys.\ Rev.\ C {\bf 89}, no. 1, 014902 (2014)
  doi:10.1103/PhysRevC.89.014902
  [arXiv:1311.5339 [nucl-th]].



\bibitem{Giacalone:2017uqx} 
  G.~Giacalone, J.~Noronha-Hostler and J.~Y.~Ollitrault,
  Phys.\ Rev.\ C {\bf 95}, no. 5, 054910 (2017)
  doi:10.1103/PhysRevC.95.054910
  [arXiv:1702.01730 [nucl-th]].



\bibitem{Albacete:2018bbv} 
  J.~L.~Albacete, P.~Guerrero-Rodríguez and C.~Marquet,
  arXiv:1808.00795 [hep-ph].



\bibitem{Aad:2013xma} 
  G.~Aad {\it et al.} [ATLAS Collaboration],
  JHEP {\bf 1311}, 183 (2013)
  doi:10.1007/JHEP11(2013)183
  [arXiv:1305.2942 [hep-ex]].



\bibitem{Voloshin:2007pc} 
  S.~A.~Voloshin, A.~M.~Poskanzer, A.~Tang and G.~Wang,
  Phys.\ Lett.\ B {\bf 659}, 537 (2008)
  doi:10.1016/j.physletb.2007.11.043
  [arXiv:0708.0800 [nucl-th]].



\bibitem{Aad:2013fja} 
  G.~Aad {\it et al.} [ATLAS Collaboration],
  Phys.\ Lett.\ B {\bf 725}, 60 (2013)
  doi:10.1016/j.physletb.2013.06.057
  [arXiv:1303.2084 [hep-ex]].



\bibitem{Chatrchyan:2013nka} 
  S.~Chatrchyan {\it et al.} [CMS Collaboration],
  Phys.\ Lett.\ B {\bf 724}, 213 (2013)
  doi:10.1016/j.physletb.2013.06.028
  [arXiv:1305.0609 [nucl-ex]].



\bibitem{Giacalone:2016eyu} 
  G.~Giacalone, L.~Yan, J.~Noronha-Hostler and J.~Y.~Ollitrault,
  Phys.\ Rev.\ C {\bf 95}, no. 1, 014913 (2017)
  doi:10.1103/PhysRevC.95.014913
  [arXiv:1608.01823 [nucl-th]].



\bibitem{Sirunyan:2017fts} 
  A.~M.~Sirunyan {\it et al.} [CMS Collaboration],
  arXiv:1711.05594 [nucl-ex].



\bibitem{Acharya:2018lmh} 
  S.~Acharya {\it et al.} [ALICE Collaboration],
  JHEP {\bf 1807}, 103 (2018)
  doi:10.1007/JHEP07(2018)103
  [arXiv:1804.02944 [nucl-ex]].



\bibitem{Mehrabpour:2018kjs} 
  H.~Mehrabpour and S.~F.~Taghavi,
  arXiv:1805.04695 [nucl-th].



\bibitem{Alver:2008zza} 
  B.~Alver {\it et al.} [PHOBOS Collaboration],
  Phys.\ Rev.\ C {\bf 77}, 014906 (2008)
  doi:10.1103/PhysRevC.77.014906
  [arXiv:0711.3724 [nucl-ex]].



\bibitem{Bhalerao:2011bp} 
  R.~S.~Bhalerao, M.~Luzum and J.~Y.~Ollitrault,
  Phys.\ Rev.\ C {\bf 84}, 054901 (2011)
  doi:10.1103/PhysRevC.84.054901
  [arXiv:1107.5485 [nucl-th]].



\bibitem{Yan:2013laa} 
  L.~Yan and J.~Y.~Ollitrault,
  Phys.\ Rev.\ Lett.\  {\bf 112}, 082301 (2014)
  doi:10.1103/PhysRevLett.112.082301
  [arXiv:1312.6555 [nucl-th]].



\bibitem{Ade:2013ydc} 
  P.~A.~R.~Ade {\it et al.} [Planck Collaboration],
  Astron.\ Astrophys.\  {\bf 571}, A24 (2014)
  doi:10.1051/0004-6361/201321554
  [arXiv:1303.5084 [astro-ph.CO]].



\bibitem{Abbasi:2017ajp} 
  N.~Abbasi, D.~Allahbakhshi, A.~Davody and S.~F.~Taghavi,
  Phys.\ Rev.\ C {\bf 98}, no. 2, 024906 (2018)
  doi:10.1103/PhysRevC.98.024906
  [arXiv:1704.06295 [nucl-th]].



\bibitem{Khachatryan:2015waa} 
  V.~Khachatryan {\it et al.} [CMS Collaboration],
  Phys.\ Rev.\ Lett.\  {\bf 115}, no. 1, 012301 (2015)
  doi:10.1103/PhysRevLett.115.012301
  [arXiv:1502.05382 [nucl-ex]].



\bibitem{Gronqvist:2016hym} 
  H.~Gr\"onqvist, J.~P.~Blaizot and J.~Y.~Ollitrault,
  Phys.\ Rev.\ C {\bf 94}, no. 3, 034905 (2016)
  doi:10.1103/PhysRevC.94.034905
  [arXiv:1604.07230 [nucl-th]].



\bibitem{Aamodt:2010pa} 
  K.~Aamodt {\it et al.} [ALICE Collaboration],
  Phys.\ Rev.\ Lett.\  {\bf 105}, 252302 (2010)
  doi:10.1103/PhysRevLett.105.252302
  [arXiv:1011.3914 [nucl-ex]].



\bibitem{Borghini:2001vi} 
  N.~Borghini, P.~M.~Dinh and J.~Y.~Ollitrault,
  Phys.\ Rev.\ C {\bf 64}, 054901 (2001)
  doi:10.1103/PhysRevC.64.054901
  [nucl-th/0105040].



\bibitem{Aad:2014vba} 
  G.~Aad {\it et al.} [ATLAS Collaboration],
  Eur.\ Phys.\ J.\ C {\bf 74}, no. 11, 3157 (2014)
  doi:10.1140/epjc/s10052-014-3157-z
  [arXiv:1408.4342 [hep-ex]].



\bibitem{Luzum:2011mm} 
  M.~Luzum,
  J.\ Phys.\ G {\bf 38}, 124026 (2011)
  doi:10.1088/0954-3899/38/12/124026
  [arXiv:1107.0592 [nucl-th]].



\bibitem{Gardim:2011xv} 
  F.~G.~Gardim, F.~Grassi, M.~Luzum and J.~Y.~Ollitrault,
  Phys.\ Rev.\ C {\bf 85}, 024908 (2012)
  doi:10.1103/PhysRevC.85.024908
  [arXiv:1111.6538 [nucl-th]].



\bibitem{Niemi:2012aj} 
  H.~Niemi, G.~S.~Denicol, H.~Holopainen and P.~Huovinen,
  Phys.\ Rev.\ C {\bf 87}, no. 5, 054901 (2013)
  doi:10.1103/PhysRevC.87.054901
  [arXiv:1212.1008 [nucl-th]].



\bibitem{Miller:2007ri} 
  M.~L.~Miller, K.~Reygers, S.~J.~Sanders and P.~Steinberg,
  Ann.\ Rev.\ Nucl.\ Part.\ Sci.\  {\bf 57}, 205 (2007)
  doi:10.1146/annurev.nucl.57.090506.123020
  [nucl-ex/0701025].


\bibitem{Wei:2018xpm} 
  D.~X.~Wei, X.~G.~Huang and L.~Yan,
  arXiv:1807.06299 [nucl-th].



\bibitem{Moreland:2014oya} 
  J.~S.~Moreland, J.~E.~Bernhard and S.~A.~Bass,
  Phys.\ Rev.\ C {\bf 92}, no. 1, 011901 (2015)
  doi:10.1103/PhysRevC.92.011901
  [arXiv:1412.4708 [nucl-th]].



\bibitem{Giacalone:2018cuy} 
  G.~Giacalone, J.~Noronha-Hostler, M.~Luzum and J.~Y.~Ollitrault,
  arXiv:1807.05557 [nucl-th].



\bibitem{Giacalone:2017dud} 
  G.~Giacalone, J.~Noronha-Hostler, M.~Luzum and J.~Y.~Ollitrault,
  Phys.\ Rev.\ C {\bf 97}, no. 3, 034904 (2018)
  doi:10.1103/PhysRevC.97.034904
  [arXiv:1711.08499 [nucl-th]].



\bibitem{Yan:2014afa} 
  L.~Yan, J.~Y.~Ollitrault and A.~M.~Poskanzer,
  Phys.\ Rev.\ C {\bf 90}, no. 2, 024903 (2014)
  doi:10.1103/PhysRevC.90.024903
  [arXiv:1405.6595 [nucl-th]].



\bibitem{Yan:2014nsa} 
  L.~Yan, J.~Y.~Ollitrault and A.~M.~Poskanzer,
  Phys.\ Lett.\ B {\bf 742}, 290 (2015)
  doi:10.1016/j.physletb.2015.01.039
  [arXiv:1408.0921 [nucl-th]].



\bibitem{Kharzeev:2004ey} 
  D.~Kharzeev,
  Phys.\ Lett.\ B {\bf 633}, 260 (2006)
  doi:10.1016/j.physletb.2005.11.075
  [hep-ph/0406125].



\bibitem{Voloshin:2004vk} 
  S.~A.~Voloshin,
  Phys.\ Rev.\ C {\bf 70}, 057901 (2004)
  doi:10.1103/PhysRevC.70.057901
  [hep-ph/0406311].



\bibitem{Bhalerao:2006tp} 
  R.~S.~Bhalerao and J.~Y.~Ollitrault,
  Phys.\ Lett.\ B {\bf 641}, 260 (2006)
  doi:10.1016/j.physletb.2006.08.055
  [nucl-th/0607009].



\bibitem{Jia:2014pza} 
  J.~Jia and S.~Radhakrishnan,
  Phys.\ Rev.\ C {\bf 92}, no. 2, 024911 (2015)
  doi:10.1103/PhysRevC.92.024911
  [arXiv:1412.4759 [nucl-ex]].



\bibitem{Hirano:2005xf} 
  T.~Hirano, U.~W.~Heinz, D.~Kharzeev, R.~Lacey and Y.~Nara,
  Phys.\ Lett.\ B {\bf 636}, 299 (2006)
  doi:10.1016/j.physletb.2006.03.060
  [nucl-th/0511046].



\bibitem{Schenke:2012wb} 
  B.~Schenke, P.~Tribedy and R.~Venugopalan,
  Phys.\ Rev.\ Lett.\  {\bf 108}, 252301 (2012)
  doi:10.1103/PhysRevLett.108.252301
  [arXiv:1202.6646 [nucl-th]].



\bibitem{Niemi:2015qia} 
  H.~Niemi, K.~J.~Eskola and R.~Paatelainen,
  Phys.\ Rev.\ C {\bf 93}, no. 2, 024907 (2016)
  doi:10.1103/PhysRevC.93.024907
  [arXiv:1505.02677 [hep-ph]].



\bibitem{Nagle:2018ybc} 
  J.~L.~Nagle and W.~A.~Zajc,
  arXiv:1808.01276 [nucl-th].



\bibitem{DiFrancesco:2016srj} 
  P.~Di Francesco, M.~Guilbaud, M.~Luzum and J.~Y.~Ollitrault,
  Phys.\ Rev.\ C {\bf 95}, no. 4, 044911 (2017)
  doi:10.1103/PhysRevC.95.044911
  [arXiv:1612.05634 [nucl-th]].


\end{thebibliography}
\end{document}